\documentclass[sigconf]{acmart}

\usepackage{booktabs}
\usepackage{algorithm}
\usepackage{algpseudocode}
\usepackage{bbold}
\usepackage{enumitem}
\usepackage{subfigure}
\usepackage{colortbl}
\usepackage{amsmath}

\usepackage{graphicx}
\usepackage[onehalfspacing]{setspace}

\newcommand{\sz}[1]{\textcolor{blue}{#1}}

\newtheoremstyle{mydef}
{2ex}
{2ex}
{\itshape}
{}
{\scshape}
{: }
{0.5em}
{}
\theoremstyle{mydef}
\newtheorem{mydef}{Definition}

\usepackage{enumitem}
\setlist[itemize,1]{leftmargin=0.4cm}

\copyrightyear{2020}
\acmYear{2020} 
\setcopyright{iw3c2w3}
\acmConference[WWW '20]{Proceedings of The Web Conference 2020}{April 20--24, 2020}{Taipei, Taiwan} 
\acmBooktitle{Proceedings of The Web Conference 2020 (WWW '20), April 20--24, 2020, Taipei, Taiwan}
\acmPrice{}
\acmDOI{10.1145/3366423.3380205}
\acmISBN{978-1-4503-7023-3/20/04}

\begin{document}




\title{Novel Entity Discovery from Web Tables}
\author{Shuo Zhang}
\affiliation{%
  \institution{Bloomberg}
  \city{London}
  \country{United Kingdom}
}
\email{szhang611@bloomberg.net}
\author{Edgar Meij}
\affiliation{%
  \institution{Bloomberg}
  \city{London}
  \country{United Kingdom}
}
\email{emeij@bloomberg.net}
\author{Krisztian Balog}
\affiliation{%
  \institution{University of Stavanger}
  \city{Stavanger}
  \country{Norway}
}
\email{krisztian.balog@uis.no}
\author{Ridho Reinanda}
\affiliation{%
  \institution{Bloomberg}
  \city{London}
  \country{United Kingdom}
}
\email{rreinanda@bloomberg.net}




\begin{abstract}
When working with any sort of knowledge base (KB) one has to make sure it is as complete and also as up-to-date as possible. 
Both tasks are non-trivial as they require recall-oriented efforts to determine which entities and relationships are missing from the KB. 
As such they require a significant amount of labor. 
Tables on the Web on the other hand are abundant and have the distinct potential to assist with these tasks. 
In particular, we can leverage the content in such tables to discover new entities, properties, and relationships.
Because web tables typically only contain raw textual content we first need to determine which cells refer to which known entities---a task we dub \emph{table-to-KB matching}. 
This first task aims to infer table semantics by linking table cells and heading columns to elements of a KB.
We propose a feature-based method and on two public test collections we demonstrate substantial improvements over the state-of-the-art in terms of precision whilst also improving recall. 
Then second task builds upon these linked entities and properties to not only identify novel ones in the same table but also to bootstrap their type and additional relationships. 
We refer to this process as \emph{novel entity discovery} and, to the best of our knowledge,
it is the first endeavor on mining the unlinked cells in web tables. 
Our method identifies not only out-of-KB (``novel'') information but also novel aliases for in-KB (``known'') entities.
When evaluated using three purpose-built test collections, we find that our proposed approaches obtain a marked improvement in terms of precision {over our baselines} whilst keeping recall stable.
\end{abstract}

\keywords{Novel entity discovery; entity linking; KBP; tabular data extraction}

\maketitle


\section{Introduction}
\label{sec:int}

Knowledge bases (KBs) are typically far from complete, up-to-date, and error-free and require sometimes elaborate methods for maintenance, often with humans in the loop. 
The Web, on the other hand, contains vast amounts of semi-structured data in the form of HTML tables found on Web pages which may serve as a unique resource to complement and update KBs. 
In particular, \emph{relational tables}---centering around a set of entities in the so-called \emph{core column}, as well as their attributes in the remaining columns---are especially amenable to automatically having their contents extracted and linked to a KB. 
A table is said to be \emph{linkable} if the KB has a matching entity for any of the mentions in its core column, i.e., the subject column that contains most entity-based mentions~\citep{Zhang:2013:MEA, Lehmberg:2016:WTC}.
Commonly, this is the leftmost column in a table and the other columns would correspond to attributes or relationships of these entities.

%
\begin{figure}[t]
	\centering
	\includegraphics[scale=0.36]{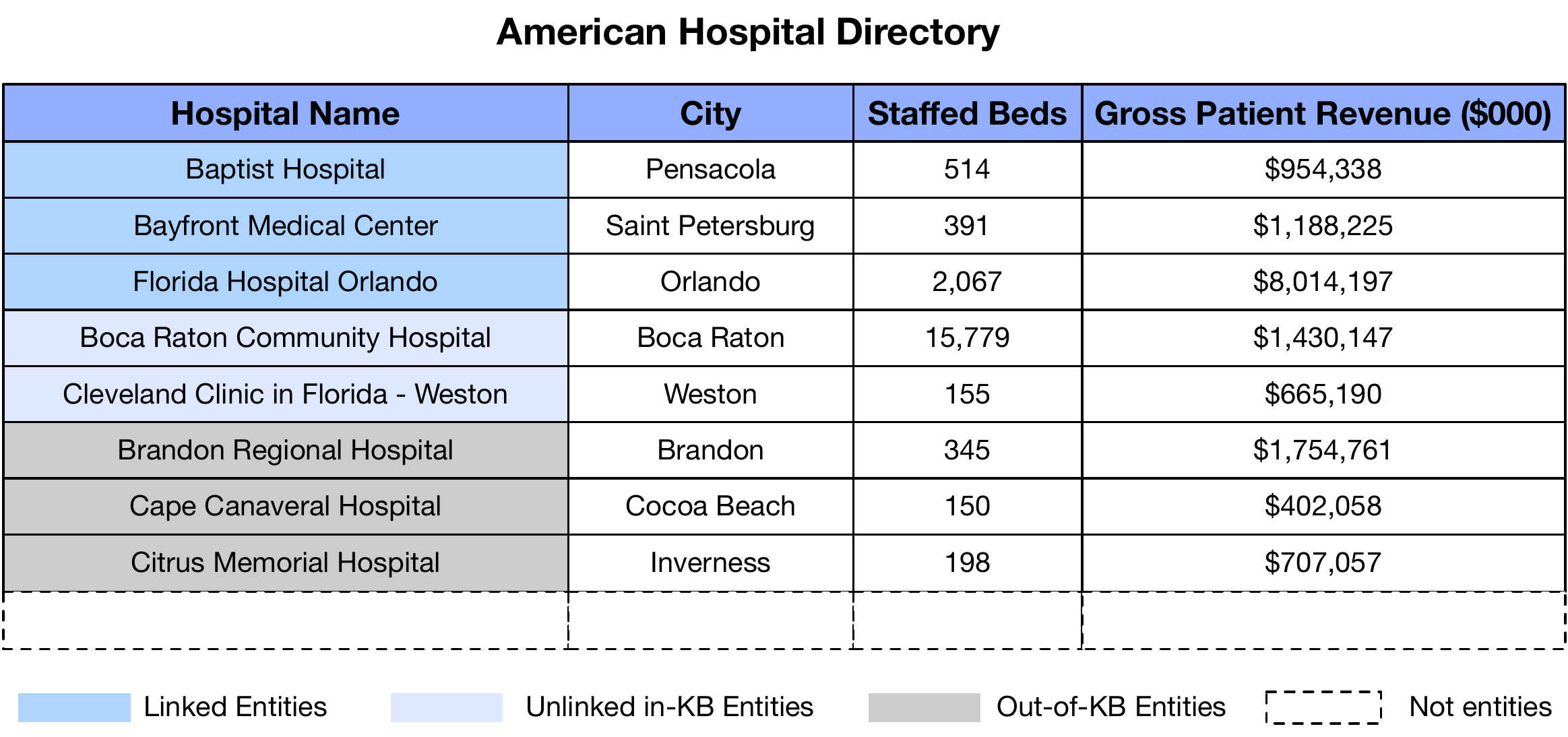}
	\caption{Example of linking entities in a Web table using a state-of-the-art method. In the core column (the leftmost column that has the most mentions identified) it links the first three rows to the KB correctly, whilst failing for the following two. The three grey rows do not exist in the KB and are potentially novel entities.}
	\label{fig:ti}
\end{figure}

In this paper we investigate to what extent we can discover new knowledge from such tables on the Web, either in the form of new entities, new properties, or new relationships. 
To this end, we define two tasks. 
The first aims to match a table with elements of the KB (DBpedia), and thus includes a step that links table cells to KB entities, a step that collectively disambiguates entities based on the dominant type of the inferred entities, as well as a step to match table headings to KB relationships. 
After the first task we find that some table cells may be left unlinked to the KB. 
The second and main task, therefore, aims to predict whether those unlinked mentions are either significant entities that do not exist in the KB but should, entities that already exist in the KB but were not linked, or are mentions that do no warrant inclusion. 

Most table-to-KB approaches only focus on ``obviously linkable'' table cells and ignore any other unlinked tabular data, and the coverage of such methods is therefore limited. 
For instance, the T2K framework by~\citet{Ritze:2016:PPW} manages to match only 2.85\% of Web tables to DBpedia with at least one correspondence. 
Given this low level of recall, we hypothesize that we may find additional relevant entities in the remaining, unlinked parts of the tables.
For instance, ~\citet{Quercini:2013:EDA} find that ``22\% of entities in tables in Google Fusion Tables
are actually represented in either Yago, DBpedia, and Freebase''. 
The out-of-KB entities include those entities that should be but have not been added to the KB yet either because they are, e.g., novel (the average latency of a Wikipedia article creation behind news appearance is 133 days~\cite{Wu:2016:EMF}) or because they may not have been notable enough to be included in the KB. 
See Fig.~\ref{fig:ti} for an illustration.

Existing methods from the literature also rely heavily on the columnar values, which are very heterogeneous and hard to normalize. 
We introduce an enhanced matching method that is able to cover more unlinked in-KB values. 
We address this by proposing a feature-based entity linking method based only on the core column mentions and find that it outperforms state-of-the-art methods from the literature using two public entity linking test collections. 

Moreover, to the best of our knowledge there is no prior work focusing on novel entity discovery from tables.
We aim to address this gap and propose two entity discovery methods as baselines and propose yet another novel method using the characteristics of the original tables, the distance between each mention and the KB, and their semantic similarities. 
This method achieves the best performance when evaluated on a purpose-built test collection.
We further resolve the mentions using novel type resolution and mention resolution methods, and achieve compelling performance.
%
In sum, this paper makes the following contributions.
%
\begin{itemize}
	\item 
	  We develop a well-performing table-to-KB matching method (Sect.~\ref{sec:tm}), which includes entity linking (Sect.~\ref{sec:sub:el}) and column heading matching (Sect.~\ref{sec:sub:cm}).
	\item 
	  We apply our table-to-KB matching method to a whole table corpus and obtain mention-entity and heading-property links for about 3 million Web tables. 
	  The linked tables represent a useful resource for future applications, and we make them publicly available.\footnote{The three test collections for novel entity discovery for Web tables, entity type and mention resolution, as well as the mention-entity and heading-property correspondences for 3M tables can be found at \url{https://doi.org/10.5281/zenodo.3627274}.}
	\item 
	  We propose a novel entity discovery task from Web tables (Sect.~\ref{sec:ned}) and introduce various methods (Sect.~\ref{sec:sub:ned}). 
	  We build an entity discovery test collection with 20K core column mentions sampled from Web tables and will also make those publicly available.
	\item 
	  Finally, we propose a very effective method for entity resolution to group mentions that refer to the same, out-of-KB entity, and infer a type for it (Sect.~\ref{sec:sub:er}).
\end{itemize}
\section{Related Work}
\label{sec:rw}

Our work is related to table-to-KB matching, novel entity discovery and KBP. As already indicated above, our methods are most similar to the state-of-the-art methods presented in~\citep{Wu:2016:EMF}, ~\citep{Efthymiou:2017:MWT}, ~\citep{Ritze:2015:MHT}, and ~\citep{Lin:2012:NNP} and we have included all of these approaches as baselines. 

\emph{Table-to-KB matching.}
Table-to-KB matching aims to annotate tables with entities and predicates.
It is a critical step for many applications like KBP~\citep{Chirigati:2016:KEU, Ritze:2015:MHT, RitzeB:2017:MWT,Ritze:2016:PPW, Dong:2014:KVW, Venetis:2011:RST, Yoones:2014:KBA, Zwicklbauer:2013:TDW}.
To match a certain Web table corpus of WDC~\citep{Lehmberg:2016:LPC} to DBpedia, ~\citet{Ritze:2015:MHT} propose the T2K match framework. They further introduce three fusion methods for knowledge base augmentation. As a follow-up, ~\citet{RitzeB:2017:MWT} focus on the utility of features extracted from tables and DBpedia for the purpose of matching them.
For tabular entity linking, \citet{Efthymiou:2017:MWT} propose two hybrid methods of two components in different orders, which are considered as state-of-the-art.
The first component compares terms shared between the table and the entity's description and its relations in the KB. 
The second is an entity embedding-based method, which uses Word2vec to capture the structure of the neighborhood of each entity in the KB. 
The trained embeddings are later used for annotating tables by considering the columns with text values.
In this paper, we perform the same tasks with \citet{RitzeB:2017:MWT}.

Table-to-KB matching is largely overlapping with table interpretation.
The goal of table interpretation is to uncover the table semantics with the help of knowledge bases and make tabular data readable by machines. It covers tasks such as column type identification~\cite{Venetis:2011:RST, Mulwad:2010:ULD, Fan:2014:AHM}, entity linking~\cite{Bhagavatula:2015:TEL, Efthymiou:2017:MWT, Mulwad:2010:ULD, Lehmberg:2016:LPC, Ibrahim:2016:MSE, Zhang:2013:ISM, Fan:2014:AHM} and relation extraction~\cite{Venetis:2011:RST, Mulwad:2010:ULD, Mulwad:2013:SMP}.
In specific, relation extraction aims to extract relationships between a pair of tabular cells with the help of column type identification and entity linking. 
Relation extraction focuses on relations or facts centering around linkable entities. 
\citet{Bhagavatula:2015:TEL} propose TabEL, which employs a graphical model to jointly model table interpretation.
Such joint models are based on a strong assumption that the relations expressed can be mapped to the target KB. 
However, less than 3\% of WDC tables can be matched to DBpedia by entity linking, and only 0.9\% of these tables having relations can be mapped to DBpedia~\citep{Ritze:2016:PPW}, i.e., the overwhelming majority of the tables we are working on in this paper do not meet this assumption.
Because of this, we do not consider TabEL as our baseline for entity linking.  
Recently, a bunch of deep learning methods for entity linking can also be performed for tables~\citep{Le:2018:IEL,Ganea:2017:DJE,Kolitsas:2018:ENE}. 
For example, \citet{Le:2018:IEL} propose an end-to-end entity linking method by treating relations between mentions as latent variables.

%
\begin{figure*}[t]
	\centering
	\includegraphics[scale=0.52]{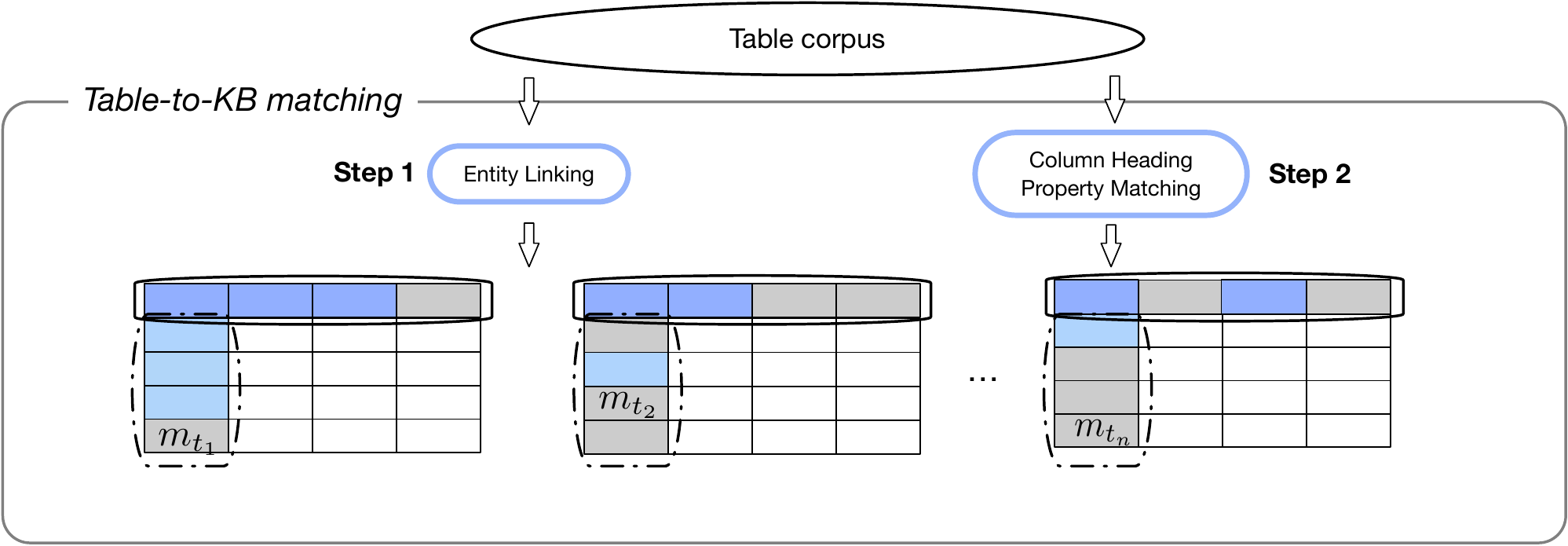}
	\caption{Illustration of table-to-KB matching. }
	\label{fig:pip1}
\end{figure*}
\begin{figure*}[t]
	\centering
	\includegraphics[scale=0.52]{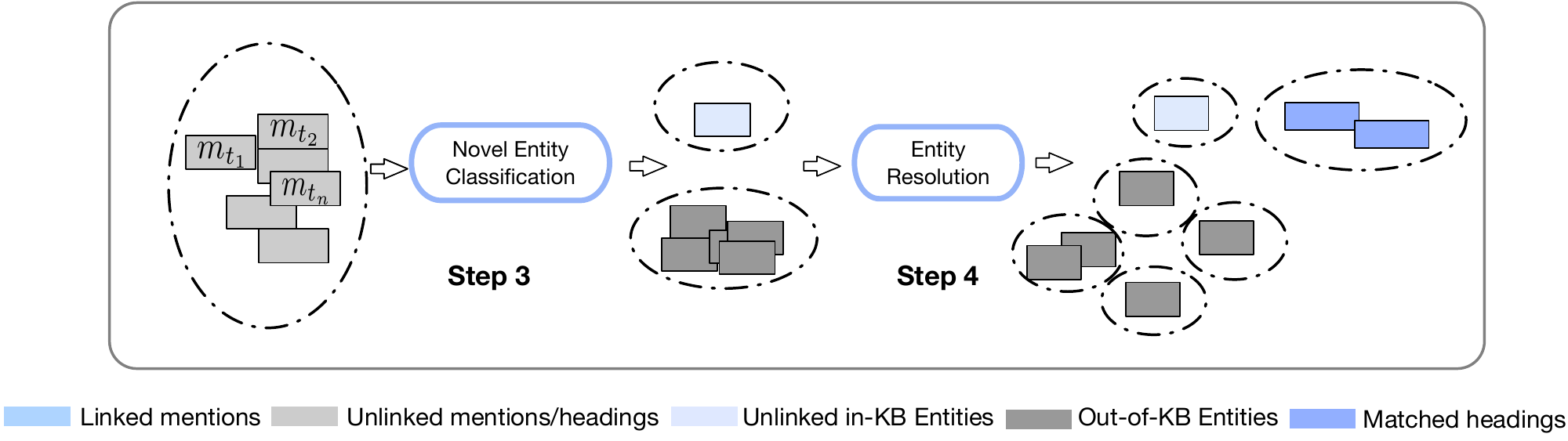}
	\caption{Illustration of novel entity discovery. }
	\label{fig:pip2}
\end{figure*}

\emph{Novel entity discovery.}
Whilst table-to-KB is well-studied, novel entity discovery is an emerging and an, as-of-yet, unsolved problem~\cite{Wu:2016:EMF}. 
~\citet{Lin:2012:NNP} introduce the task of unlinkable noun phrase problem to predict if a noun phrase is an entity, as well as its fine-grained semantic types. 
Focusing on determining whether any noun phrase is an entity and what semantic types it holds, \citet{Lin:2012:NNP} train a classifier with features primarily derived from a timestamped corpus. 
The main intuition is that usage patterns differ across time between in-KB entities and out-of-KB entities and they operationalize this idea by computing the best fit {line (least-squares regression )} for usage over time.
The semantic type is determined by observing the types of linked entities with the same textual relations.
~\citet{Wu:2016:EMF} propose a method to discover novel entities in news and Web data by exploring multiple feature spaces, including context, neural embedding, topical, query and lexical spaces.
~\citet{David:2018:TBC} track entities that emerge in public discourse to gain insights into how these are added to Wikipedia.
The above methods leverage unstructured text. As far as we know, we are the first to discover new entities from tables.

\emph{KBP using tables.}
KBP using tables aims to find new facts or relations for completing the KBs.
Most studies on KBP using tables focus on matching tables to the knowledge base such as~\citep{Chirigati:2016:KEU, Ritze:2015:MHT, RitzeB:2017:MWT,Ritze:2016:PPW, Dong:2014:KVW, Venetis:2011:RST, Yoones:2014:KBA, Zwicklbauer:2013:TDW}, as we have discussed in table-to-KB matching.
\citet{Ritze:2016:PPW} work on investigating tabular data are potentially useful for KBP, and this study is a pre-step towards end-to-end knowledge base population.
\citet{Yoones:2014:KBA} propose a probabilistic method for finding new relations by collecting sentences containing pairs of entities in the same row in a table. Their method extracts the patterns around the co-occuring entities and estimate the probability of possible relations that can be added to the repository. They only focus on entity pairs, in contrast with \citet{Cannaviccio:2018:LWT} who do not limit their work to entities. They leverage the table schema and, similar to~\cite{Yoones:2014:KBA}, they focus on refining relations.
Novel entity discovery also relies on table-to-KB matching. 
However, the main difference between KBP using tables and novel entity discovery is that the former only uses ``obviously linkable'' table cells and ignores any unlinked tabular data. 

\emph{Summary of differences.}
There are several perspectives to distinct our work from the existing methods. 
\emph{KBP using tables} and \emph{novel entity discovery} rely on table-to-KB matching tasks like entity linking and schema matching. 
However,
the current methods are limited to the coverage problem.
We, therefore, aim to improve the matching of tables to KBs further.
Additionally, most of these tasks focus on the linked data and simply ignore the unlinked tabular data.
Novel entity discovery dedicates to the unlinked tabular data.
Lastly, existing novel entity discovery only considers text data while tabular data is an under-explored source. 
The discovered entities will enhance the performance of table-related tasks like KBP.



\vspace*{-0.5\baselineskip}
\section{Problem Statement  and Overview}
\label{sec:pso}

In this section, we give a formal description of the tasks.
%
%
We assume the presence of a single core column in the relational tables like most existing work~\citep{Zhang:2018:AHT,Ritze:2016:PPW, Chirigati:2016:KEU,Venetis:2011:RST, Zhang:2019:ADC}. 
For any relational table $t= (M, H, V)$ as found on the Web, where $M=<m_1, m_2, ..., m_l>$ is a sequence of core column mentions, $H=<h_1, h_2,..., h_n>$ is a sequence of heading labels, and $V$ is the matrix of the remaining tabular content, we carry out \emph{table-to-KB matching} (Sect.~\ref{sec:tm}) which includes entity linking (Sect.~\ref{sec:sub:el}) and column heading property matching (Sect.~\ref{sec:sub:cm}). See Fig.~\ref{fig:pip1} as an illustration.
Subsequently we perform \emph{novel entity discovery} based on this information (Sect.~\ref{sec:ned}), which includes novel entity classification (Sect.~\ref{sec:sub:ned}) and entity resolution (Sect.~\ref{sec:sub:er}). See Fig.~\ref{fig:pip2} as an illustration.

\begin{mydef}[Table-to-KB matching]
	Table-to-KB matching is the task of {linking table cells} to entities and predicates in a target KB.
\end{mydef}

\begin{mydef}[Novel entity discovery]
	Novel entity discovery is the task of predicting the unlinked mentions in a relational table as in-KB, out-of-KB, or not entities.
\end{mydef}

There is a tradeoff between table-to-KB matching and novel entity discovery. 
I.e., if we manage to achieve a high matching coverage, the unlinked tabular data are most likely all novel entities. 
To keep track of the performance of each component, we address them in a pipeline architecture instead of end-to-end.
The first task, table-to-KB matching, thus includes steps 1 and 2 below, whereas novel entity discovery starts from step 3.
\begin{description}
	\item[\textbf{Step 1:} {Entity Linking.}] 
	  This step aims to link the core column mentions $M$ to the knowledge base for relational tables, i.e., given a collection of tables, we find the possible mention-entity correspondences for linkable tables.
	\item[\textbf{Step 2:} {Column Heading Property Matching.}]
	  We then aim to match the table headings $H$ (including core column heading) to the KB properties for all the linkable tables with the help of entity linking. 
	\item[\textbf{Step 3:} {Novel Entity Classification.}]
	  We take every unlinked mention $m$ in the core columns from the linkable tables and identify if they are entities that should be included in the KB, i.e., we classify them as in-KB, out-of-KB, or not an entity. 
	\item[\textbf{Step 4:} {Entity Resolution.}]
	  We then group mentions that refer to the same entity and assign the same type. 
	  In the end, the unlinked in-KB mentions are taken as additional surface forms for those entities, while the out-of-KB entities are returned as new KB entries. 
\end{description}


\section{Table-To-KB Matching Across Tables}
\label{sec:tm}
In this section, we address the problem of table-to-KB matching, which aims to identify which table cells refer to already known entities in the KB. As such, this task includes entity linking and column heading property matching. 
We first execute entity linking for all the tables (Sect.~\ref{sec:sub:el}), and then the linked mentions in the linkable tables will help to match the column headings to properties in the KB (Sect.~\ref{sec:sub:cm}). 
The way we disambiguate mentions to entities leverages the information that is specific to a table and, as such, the same mention in different tables might be linked differently. 

\subsection{Entity Linking}
\label{sec:sub:el}
Entity linking for tables aims to identify relevant entities for table cells and, as a result, identify all \emph{linkable} tables, i.e., those tables with at least one mention-entity correspondence.
Intuitively, core column entities tend to be of similar type~\cite{Ritze:2015:MHT}. 
Candidate selection is important to maintain reasonable coverage.
A very small portion of columnar data can be mapped to the target KB and we therefore do not leverage that for entity linking.
{Based on the above}, we propose a novel feature-based classification method for entity linking on Web tables {that incorporates the table type as well as lexical and semantic similarities.} 
Our method has three steps.
Candidate selection aims to find candidate entities for each mention, 
candidate classification predicts if a mention can be linked to a candidate entity, and 
entity disambiguation makes sure one mention at most can be linked to one candidate.

\subsubsection{Candidate Selection}
\label{sec:el:cs}
For a single relational Web table $t$, we take the sequence of core column mentions $M$ in $t$ as input. 
For each $m \in M$, we find the top-$k$ highest-scoring candidates using a traditional retrieval method. Specifically, we use the Wikipedia search {API} 
which combines a keyword-based method 
with article popularity, and keep them as $C$.
$C$ is a $l \times k$ matrix, where $c_{ij}$ ($1\leq i \leq l, 1\leq j\leq k$) is the $i_{th}$ mention's $j_{th}$ candidate.
We then determine the table type by taking a majority vote ($\mathbb{M}$) among the KB types of the top-$k$ results, i.e.,
\begin{equation*}
	y_t = \mathbb{M}\{y_c | c \in C_{[:1]}\},
\end{equation*}
where $y_c$ denotes the KB {types} of a candidate $c \in C$.
\citet{Ritze:2016:PPW} exclude the candidates with a different type in this step, which potentially results in a low recall.
We therefore keep all the candidates and consider $y_c$ a ``soft'' constraint.

\subsubsection{Candidate Classification}
\label{sec:el:cc}
We extract two types of features and train a classifier to predict if $m$ is possibly linkable to a candidate entity. I.e., for any mention $m_i$ and its candidate $c_j$ in $C_{[i:]}$:
\begin{equation}
	C'_{[i:j]} = \mathbb{1}(m_i, c_j).
\end{equation}
$C'$ is a binary matrix and $C'_{[i:j]}=1$ indicates $m_i$ can be linked to $c_j$, not otherwise. The two types of features are as follows.
\begin{table}
	\caption{Lexical similarities. We use $LD$ to denote levenshtein distance, and $L$ and $W$ to denote the character and term sets.}
	\begin{tabular}{l l}
		\toprule
		Type & Expression \\
		\toprule
		Edit Distance & $LD(m,c)/ \mathrm{max}\{|m|, |c|\}$ \\
		Letter Distance & $|L_m \cap L_c| / \mathrm{max}\{|m|, |m|\}$ \\
		Jaccard Similarity & $|W_m \cap W_c| / |W_m \cup W_c| $\\
		Substring Indicator & $\mathbb{1}$ if $m \in c$ or $c \in m$ else $\mathbb{0}$\\
		\bottomrule
	\end{tabular}
	\label{tbl:ls}
\end{table}
\begin{itemize}
	\item \textbf{Lexical Similarity.} The first group of features includes four types of lexical similarity, which are listed in Table~\ref{tbl:ls}.
We first adapt a normalized Levenshtein distance between $m$ and $c$.
Similarity at the character level aims to address spelling mistakes.
For instance, `Cisco Teechnology, Inc.'', ``Cisco Technologiy, Inc.'', and ``Cisco Technolgy, Inc.'' are actual mentions found in Web tables.
We compute the Jaccard similarity based on the terms.
We use a binary indicator to signify whether one is a substring of the other. 
	\item \textbf{Semantic Simlarity.} Motivated by~\citet{Efthymiou:2017:MWT}, we additionally consider three semantic features.
We employ a deep semantic matching method, which is an enhancement of DRMM~\citep{Guo:2016:DRM} for short text and has proven effective to generate tables~\citep{Zhang:2018:OTG}. We instantiate this matching model with three matching pairs: (i) $\phi(m, c)$, (ii) $\phi(m+y_t, c+y_c)$, and (iii) $\phi(m, d_c)$, where $d_c$ is the candidate entities' textual description, and $\phi$ is the matching score.
Besides these, we also use the rank of the Wikipedia search, the binary type indicator that indicates if $y_{c}$ exists, and another binary indicator to show if $y_{c}$ is the same as $y_t$.
We also consider a binary indicator to identify if a disambiguation tag (like ``(film)'') is found in the title of $c$.
\end{itemize}

\subsubsection{Entity Disambiguation}
All the above features are used to predict whether $m_i$ can be linked to a candidate $c$ in $C_{[i:]}$, by calculating the binary matrix $C'$, which corresponds to $C$ and $c'_{ij}=1$ indicates $c_j$ is a possible entity for $m_i$.
Note that this method for entity linking might link a mention to multiple entities in the KB, i.e., there might possibly be multiple 1s in $C'_{[i:]}$. 
We implement the following filter to ensure that any mention will be linked to a single entity at most. We utilize table type as a ``lightweight'' disambiguator and our filter will keep the top-ranked candidate having the same type with the table.
After this step, $m$ can at most be linked to one entity.

\subsection{Column Heading Property Matching}
\label{sec:sub:cm}

Once we have identified relevant entities, we apply column matching to link column headings to the KB properties.
T2K Match~\cite{Ritze:2016:PPW} provides a good starting point for column matching. However, T2K only utilizes the column values. 
To additionally investigate how heading label similarity works for this task,
we develop a novel feature-based method utilizing both facts and label similarity. 

We dub our method ``Entity-assisted Column Matching'' as it leverages the found entity links $L$ using the method in Sect.~\ref{sec:sub:el} to improve matching the column headings with the KB properties. 
Following~\cite{Ritze:2015:MHT}, for each heading $h$ in $H$, we take all the properties of $L$ in the KB as the candidates, i.e., $P=\{p | p \in \{<e_{m_l}, p, o> | m_l \in L\}\}$, where $\{<e_{m_l}, p, o> | m_l \in L\}$ are the corresponding triples of all related entities of $L$ in the KB.
To distinguish if a candidate property is matched to $h$, we develop a feature-based binary classifier by utilizing the following features.
\begin{itemize}
	\item \textbf{Naive Features.} Features include an indicator identifying if the column is the core column as well as the lengths of the heading and predicate.
	\item \textbf{Label Similarity.} T2K Match does not consider the label similarity, but heading labels actually carry information. 
For instance, the top-5 most popular headings in the WDC table corpus according to~\cite{Ritze:2015:MHT} are ``releaseDate'', ``elevation'', ``populationTotal'', ``location'', and ``industry'' respectively; all with the exact same ontological predicates in the KB.
We compute the label similarity between $h$ and $p$ using the same four string level similarity methods in entity linking, i.e., edit distance, letter distance, Jaccard similarity, and substring indicator (cf. Table~\ref{tbl:ls}).
	\item \textbf{Value Similarity.}  
Following~\cite{Ritze:2015:MHT}, we use $V^L_h$ to denote the table values of $L$ in the column of $h$, and $V^L_p$ to denote the KB values/objectives of $L$ for property $p$. 
The KB values and tabular values in the column are classified as time, numerical, string, or other data types. 
We type ambiguous values, such as ``1836'', as both ``numerical'' and ``time''.
We assign a type to $V^L_h$ as column data type by majority voting and properties holding the same data type with the $V^L_h$ are kept as the final candidates.
$V^L_h$ with the values in the KB. 
We aggregate all pairwise value similarities (PVS) between $V^L_h$ and $V^L_p$ {as features}, i.e.,
\begin{equation}
  PVS = \digamma(\{sim(v^L_h, v^L_p)| v^L_h \in V^L_h, v^L_p \in V^L_p\}),
\end{equation}
where $\digamma$ includes $max$, $sum$ and $avg$ and $sim$ is computed based on the data type. 
Specifically, we compute the standard deviation for time (in years) and for (normalized) numbers.
We use edit distance to compute the similarity of strings. 
\end{itemize}
%
These ten features are used to predict whether $p$ can be linked to $h$.

\section{Novel Entity Discovery}
\label{sec:ned}

With the help of entity linking in Sect.~\ref{sec:sub:el} and column heading matching in Sect.~\ref{sec:sub:cm}, 
we turn to determine which unlinked mentions might refer to novel entities and, moreover, which of these mentions in disparate tables refer to the same, hitherto unseen entity. 
As far as we know, there is no prior work on novel entity discovery from tables, and we deem this task as one main contribution.
Concretely, we classify the unlinked mentions in all the linkable tables using an entity discovery classifier (Sect.~\ref{sec:sub:ned}).
We then resolve the mentions that refer to the same entities in Sect.~\ref{sec:sub:er}. 

\subsection{Novel Entity Discovery Classification}
\label{sec:sub:ned}

In Sect.~\ref{sec:sub:el} we linked core column mentions in relational tables to entities in the KB. 
We aim to classify any remaining, unlinked mentions as in-KB, out-of-KB, or not entities.
In-KB entities refer to those that already exist in the KB, but for which the entity linker fails. 
Out-of-KB entities are novel entities that we aim to discover. 
The fact that they do not exist in the KB yet may be a result of them either not being prominent enough~\cite{Lin:2012:NNP} or not timely enough~\cite{Wu:2016:EMF}.
Not entities are mentions such as ``sjksjjkjjadk'' that we regard as noise. 
We leave correcting and/or linking these for future work.

We assume that the characteristics and the semantic distance and similarity between the mentions and the KB can help to identify if a mention refers to an entity. Under this intuition, we introduce three types of methods based on the table characteristics and the similarity between tables and the KB, and combine them into a novel model.

\subsubsection{Origin Characteristic}

For an unlinked mention $m$, we define the linkable tables $T_m$ which have $m$ in their core columns as the \emph{origin tables} of $m$. 
We assume that the characteristics of the origin tables help to classify any unlinked mentions, e.g., if a mention appears in tables in which the the majority of mentions are linkable, this mention will have a higher chance of being an entity as well, and vice versa. 
Following this intuition, we propose a number of methods that leverage multiple tables $T_m$ for identifying novel entities for the unlinked mentions. 
We consider the number of tables having $m$ in the core column ($|T_m|$) and the number of identical core columns having $m$, i.e., tables having the same set of mentions in the core column are treated as a single table which happens, e.g., when tables are reused over time on a certain website. 
We use a binary indicator to identify whether a mention is the same as the header in the column.

For table $t \in T_m$, we consider two properties: (i) the number of linked mentions $n^l_t$ and (ii) the linking rate $r^l_t$ which is defined as:
\begin{equation}
  r^l_t = \frac{n^l}{|M_t|},
\end{equation}
where $M_t$ is the sequence of mentions in the core column of table $t$. 
We use $N^l_t = [n^l_t | t \in T_m]$ and $R^l_t = [r^l_t | t\in T_m]$ to denote the feature collections of $T_m$ regarding to the number of linked mentions and link ratio. 
We further consider the aggregation methods to get two types of features, i.e., $\digamma (N^l_t)$ and $\digamma (R^l_t)$, where $\digamma$ includes $sum$, $max$, $min$, $avg$ and $std.$ $deviation$ respectively. 
Additionally, we utilize the matched columns from Sect.~\ref{sec:sub:cm}, i.e., we aggregate the number of matched headings of $T_m$ as above.

\subsubsection{Saliency-assisted Entity Discovery}
\begin{figure}[t]
\vspace*{-0.5\baselineskip}
	\centering
	\includegraphics[scale=0.45]{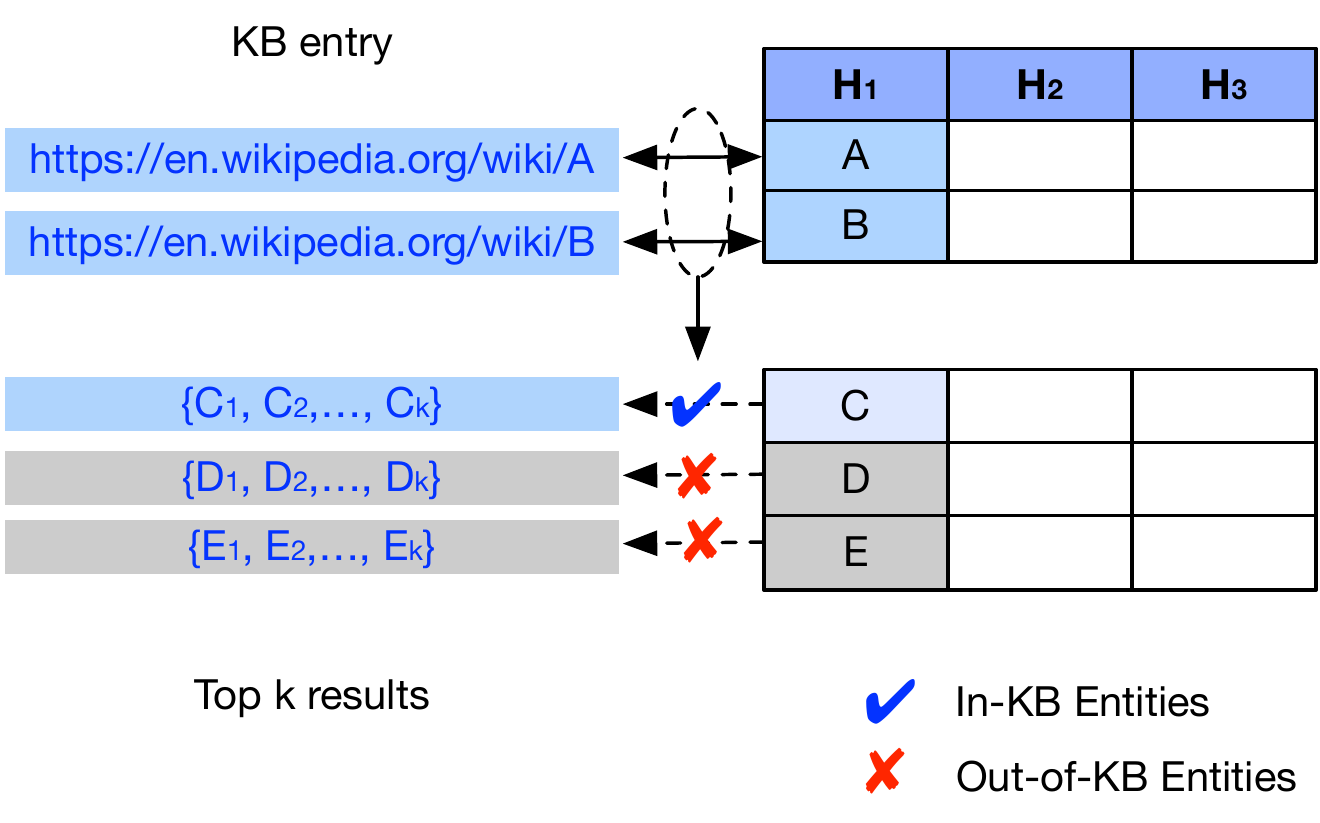}
	\vspace*{-0.5\baselineskip}
	\caption{Illustration of salient inference entity discovery.}
	\label{fig:emc}
\end{figure}
Inspired by the idea proposed by~\citet{Wu:2016:EMF}, where in-KB entities appear in the top-$k$ ranked candidates instead of just the single highest-ranked candidate, we propose a salient inference discovery method, which utilizes the linked mentions as ``inference''.
Given an unlinked mention $m$, we get all the linkable tables $T_m$ that have $m$ in their core columns. 
We take all the linked mention-entry (entity identifier) pairs ($L$) from $T_m$, i.e.,  $\{(m^l_{t}, e_{m^l_{t}})| t \in T_m\}$, and compute four pairwise label similarities, i.e., $\{\textrm{sim}(m^l_{t}, e_{m^l_{t}})\}$. 
Then we aggregate the pairwise similarities with $\digamma$, i.e., $max$, $sum$, $avg$ and $min$, to compute Mention-Entity Distance (denoted as $MED$). 
We use $MED$ to denote this salience characteristic and it is used to guide entity classification as illustrated in Fig.~\ref{fig:emc}:
\begin{equation}
  MED = \digamma(\{\textrm{sim}(m^l_{t}, e_{m^l_{t}})|t \in T_m\}).
\end{equation}
We further propose a number of features for computing the ``nearest Wikipedia distance'', denoted as $WD$. 
Specifically, we search $m$ in Wikipedia (cf. candidate selection in Sect.~\ref{sec:sub:el})
to get the top-$k$ candidates $C_m$ as an intermediate representation. 
We compute $WD$ between mention $m$ and all the candidates, and take the maximum similarity as the nearest distance, i.e., 
\begin{equation}
  WD = \textrm{max}\{\textrm{sim}(m, c_{m})| t \in C_m\},
\label{eq:wd}
\end{equation}

\subsubsection{Semantic-based Entity Discovery}
To learn a binary decision function for detecting if a mention is a novel entity, \citet{Wu:2016:EMF} define a number of semantic spaces:
\begin{equation}
  k_S = g(k_1(e_m, e_m^{top}), ..., k_n(e_m, e_m^{top})),
\end{equation}
where $e_m$ is the semantic space of mentions and $e_m^{top}$ is the semantic space of the best entity candidate, i.e., the feature space of a mention's top candidate.
$k_i (e_m, e_m^{top})$ is treated as a single feature and $g$ is the function learned from training data. 
To apply this method to the tabular mentions in our setting, we consider three features spaces: a neural embedding space, a topical space and a lexical space.  

\begin{itemize}
	\item \textbf{Neural Embedding}:
	  The contextual space of mention $m$ and $e$ are mapped into the embedding space {by replacing each term with} Word2vec ({pretrained using Google News}), and the cosine similarity between $vec(m)$ and $vec(e)$ is used as a feature.
	\item \textbf{Topical Space}:
	  An entity would occur in documents of a particular topical distribution. 
	  To consider the global topical coherence between mentions and entities, we use the entity types of the linked mentions from the tables that $m$ co-occurs with to represent the topical space of $e$, and the same type system to represent the candidate entity of an unlinked mention. 
	  The cosine similarity between those {type} representations is used to determine the topical space similarity.
	\item \textbf{Lexical Space}:
	The normalized Levenshtein distance (cf. Tbl.~\ref{tbl:ls}) between a mention and an entity name.
\end{itemize}

In the end, all three methods are represented as a set of features which we use to classify whether a mention is a novel entity or not. At the end of this step, all unlinked mentions are classified as in-KB entities or novel entities.

\subsection{Entity Resolution}
\label{sec:sub:er}

A core column mention appearing in different tables might refer to different entities, e.g., ``DSP'' can stand for ``digital signal processing'' or ``DSP Media'' company and we will need to disambiguate between entities. 
Similar mentions can also refer to the same entity when it has different surface forms, e.g., ``Cisco systems Inc.'' and ``Cisco'' might refer to the same company.
We define the former task as type resolution and the latter as surface form resolution.
Entity resolution aims to cluster the mentions from different tables that refer to the same entity and with the same type. The clustered mentions are taken as the surface forms of an entity, to be added to the KB along with any properties matched against the table headings.

\subsubsection{Type Resolution}

We assume that most of the mentions in the same column have the same type. 
Furthermore, similar mentions in similar tables can potentially share the same entity type.
For type-based mention clustering, we treat tables having the exact set of core column mentions as the same table because most of these tables only differ at the editing time, i.e., tables are reused over time. 
For an unlinked mention appearing in any pair of tables, $m_{t_1}$ and $m_{t_2}$ ($m$ appears in two different tables $t_1$ and $t_2$), we cluster them based on the types of $t_1$ and $t_2$. 
 
We use $y_{e_{m^l_t}}$ to denote the type of the linked mention $L_t$, which co-occurs with $m$ in $t_i$, and leverage the hierarchical type system in the KB, e.g., the {types} for ``London'' are {``Location'', ``PopulatedPlace'', ``Place'', ``Settlement'', ``Town'', etc.} 
We aggregate all the types of the linked mentions in $t_i$ to create a distribution over types that represents the table type, denoted as $H_{t_i}$. 
We cluster $m_{t_1}$ and $m_{t_2}$ based on cosine similarity score between $H_{t_1}$ and $H_{t_2}$. 
I.e., when $\textrm{cosine}(H_{t_1}, H_{t_2}) \geq \theta$ with $\theta$ being a threshold, $m_{t_1}$ and $m_{t_2}$ are split into two different mentions, or clustered together as being the same entity otherwise.

\subsubsection{Surface Form Resolution}

Mentions that are expressed similarly, or that are from similar tables, potentially have the same meaning.
For surface form resolution, given any pair of unlinked mentions $m_1$ and $m_2$ (where $T_1$ denotes the tables having $m_1$ and $T_2$ denotes the tables having $m_2$), we resolve them based on $t_1$ and $t_2$ (where $t_1$ is a table from $T_1$ and $t_2$ is a table from $T_2$) and the similarities between $m_1$ and $m_2$.
As mentions can be similar to each other both semantically and lexically,
we first use Word2vec to train Mention2vec embeddings by taking a sequence of mentions in a core column as a sentence and compute two mentions' cosine similarities. 
Mentions are resolved based on a similarity threshold. 
Using the four string-based similarities (cf.  Sect.~\ref{sec:sub:el}), we employ a surface form similarity method. 
Additionally, we compute the table {similarities} by different fields of headings, linked entities, table caption, page title and the text around the table as features by taking the Jaccard similarity between heading, table caption, page title, and table text for $t_1$ and $t_2$. 
We compute the cosine similarity of $H_{t_1}$ and $H_{t_t}$ and entity Jaccard overlap.
We adapt the attribute-based entity resolution in~\cite{Bhattacharya:2007:CER} into another heading similarity method by creating a similarity matrix between the heading terms from $t_1$ and $t_2$ and obtain a heading subgraph by solving the \emph{maximum weighted bipartite sub-graph problem}~\cite{Marie:2007:OSM}, which is the heading matching method from~\cite{Lehmberg:2015:MSJ}.
We further combine the string similarities and table similarities by taking them as features in order to predict if two mentions are referring to the same entity.



\vspace*{-0.5\baselineskip}
\section{Data Sources}
\label{sec:sub:ds}

In order to evaluate our approach for discovering novel entities we make use of an established dataset in the form of the WDC Web Table Corpus 2015; a large, publicly available collection of 233M relational HTML tables extracted from the July 2015 Common Crawl~\citep{Lehmberg:2016:LPC}. 
From this we use the English Relational Subset with 50.8M tables.\footnote{http://webdatacommons.org/webtables/2015/EnglishStatistics.html}
Each table comes with metadata in the form of the HTML page title, table caption, table orientation, header row location, the core column index, etc. 
Additional contextual data includes the text surrounding the table and timestamps indicating when the tables were created or edited.
%
As our knowledge base we use a DBpedia dump from around the same timeframe and restrict 
our set of entities to those that have at least one DBpedia type, yielding 4.77M entities in total. 
%
In order to obtain additional textual signal we use the corresponding Wikipedia dump with metadata.\footnote{https://dumps.wikimedia.org/other/cirrussearch/} 
%
We further filter the English relational tables as only a part of them can be linked to a KB~\cite{Ritze:2016:PPW} and retain only those that contain at least one DBpedia surface form, resulting in 16.2M tables.


\vspace*{-0.5\baselineskip}
\section{Evaluating Table-to-KB Matching}
\label{sec:tc}

Our evaluation consists of two main parts, one for identifying novel entities (cf. Section~\ref{sec:ned} and Fig~\ref{fig:pip2}) and one for evaluating the first two steps, i.e., entity linking and column heading matching as depicted in Fig~\ref{fig:pip1}. 
In this section we focus on the latter and we start by introducing our gold standard data, experimental setup, and evaluation metrics.


\subsection{Entity Linking}
\label{sec:eval:el}

For this part, we aim to evaluate linking the core column mentions to the knowledge base.


\subsubsection{Experimental setup}

\begin{table*}[t]
	\caption{Results of Entity Linking, evaluated using T2D and W2D. The two test collections differ in terms of the table size (number of rows), where the median for T2D is 100 and 21 for W2D. T2K performs differently on these collections; the other methods achieve more balanced results than T2K.}
	\begin{tabular}{l l l l l l l l}
		\toprule
		& \multicolumn{3}{c}{\textbf{T2D}} & \multicolumn{3}{c}{\textbf{W2D}} \\
		\cline{2-4} \cline{6-8}
		Method & Precision & Recall & F1 && Precision & Recall & F1 \\
		\toprule
		DB Lookup & 0.79 & 0.73 & 0.76 && 0.79 & 0.73 & 0.76\\
		T2K~\cite{Ritze:2015:MHT} & 0.90 & 0.76 & 0.82 && 0.70 & 0.63 & 0.66 \\
		Hybrid I~\cite{Efthymiou:2017:MWT} & 0.87 & 0.83 & 0.85 && 0.84 & 0.79 & 0.81 \\ 
		Hybrid II~\cite{Efthymiou:2017:MWT} & 0.85 & 0.81 & 0.83 && 0.84 & 0.79 & 0.82 \\
		\citet{Le:2018:IEL} & 0.91 & 0.83 & 0.87 && 0.81 & 0.83 & 0.82 \\
		\midrule
		Wikipedia search & 0.91 & 0.82 & 0.86 && 0.77 & 0.82 & 0.80 \\		
		Wikipedia search + Entity Disambiguation & 0.91 & 0.85 & 0.88 && 0.77 & 0.81 & 0.79 \\
		Rerank+Entity Disambiguation & \textbf{0.94} & \textbf{0.86} & \textbf{0.90} && \textbf{0.96} & \textbf{0.89} & \textbf{0.93} \\
		\bottomrule
	\end{tabular}
	\label{tbl:el}
\end{table*}
We perform entity linking for the whole WDC Table Corpus and we rely on two existing resources for entity linking evaluation on this collection.

\paragraph{Test collection 1 (T2D)} 
We use the T2Dv2 Gold Standard which consists of manually annotated row-to-entity, heading-to-property, and table-to-type correspondences of 779 {WDC} tables.\footnote{http://webdatacommons.org/webtables/goldstandardV2.html}  
Among these tables, 237 linkable tables have at least one row-to-entity correspondence and the remainder are negative samples. 
In total, there are 25,119 mention-to-entity correspondences as ground truth\footnote{The T2D dataset is arguable more simple than bigger and more complex datasets such as those considered in the Semantic Web Challenge on Tabular Data to Knowledge Graph Matching. We leave evaluation on those for future work}.

\paragraph{Test collection 2 (W2D)} 
We use 
the entity linking collection in~\cite{Efthymiou:2017:MWT}, consisting of 296 manually annotated Wikipedia tables where single rows are mapped to entities.\footnote{https://figshare.com/articles/Evaluating\_Web\_Table\_Annotation\_Methods\_From\_ \\ Entity\_Lookups\_to\_Entity\_Embeddings/5229847} 
In total, there are 5278 mention-to-entity correspondences for this collection.

We develop the entity linking model by utilizing the above two collections and perform
entity linking for tables that have a majority-voted table type, finding 3M linkable tables that have at least one mention-entity correspondence, with 1.4M mention-entity correspondences in total. 
We utilize these mention-entity correspondences and the DBpedia surface forms to resolve unlinked mentions from these linkable tables. 
The most frequently occurring linked entities are ``IBM'' and ``Microsoft'' while the most frequently occurring unlinked mentions are ``clear'' and ``2nd Day''. 
The three most frequent types are ``country'', ``athlete'' and ``settlement''.
The deep match features in Sect.~\ref{sec:el:cc} are implemented based on Matchzoo~\citep{Guo:2019:MLP}.
For evaluation, we report the results based on the two test collections. 
We split each test collection into training (80\%) and test set (20\%). 
We train the model using Random Forest and tune parameters using 5-fold cross validation on the training set. 
The results are reported solely based on the test collection that has remained completely unseen during training. 
We use Precision, Recall, and F-1 at the macro level as our evaluation metrics.

Linkable data in one table might be not linkable in another table because of the heterogeneity of tables, resulting in different contexts and thus different results for the same surface form.
The mention-entity and heading-property linkages that we obtain using the entity linking and heading property matching methods above might therefore yield different results for different tables.
For any unlinked mention $m$, we identify exact matches in the linked mentions and compare $y_m$ and table type $y_t$ and we link $m$ to the same entity if the tables have the same type. 
Additionally, we remove mentions that are obviously not named entities including numbers, dates, and email address using a set of regular expressions. 
In the end, we have a collection of linked mentions, headings, and unlinked mentions of which the last are to be classified in the next step.

\subsubsection{Baselines}
T2K Match~\cite{Ritze:2015:MHT} is an iterative matching algorithm which combines column heading matching and entity linking. 
The ``row-to-entities step'' aims to match the mentions in the core column to entities in a KB. 
Entity linking and column heading matching are iteratively reinforcing each other and the method terminates when there is convergence, i.e., when the similarity scores are not changing anymore.

We also consider three state-of-the-art methods from~\citep{Efthymiou:2017:MWT} as additional baselines.
The first method utilizes the DBpedia lookup service, i.e., it sends each mention as a query and takes the first returned entity as its corresponding entity link.\footnote{http://wiki.dbpedia.org/lookup/}  
The second and the third methods are hybrid methods of two components in different orders. 
The first component compares terms shared between the table and the entity's description and its relations in the KB. 
The second is an entity embedding-based method, which uses Word2vec to capture the structure of the neighborhood of each entity in the KB, i.e., it generates a text document by performing a random walk over the neighborhood of each entity in the KB, which is the input to Word2vec. 
The trained embeddings are later used for annotating tables by considering the columns with text values. 

We consider one additional method for entity linking based on deep learning. 
\citet{Le:2018:IEL} exploit relations between textual mentions in a document to decide if the linking decisions are compatible.
We consider the core column mentions relations in tables.
For each mention, we use the Wikipedia search API to find the top-$k$ candidate entities. We explore a range values for $k$ including 1, 5, 10, and 50. We settle on $k$=10 which attains a recall of over 98\% on our collections, and keep the same setting for candidate selection in our method.
We set d=300 and use GloVe word embedding and Wikipedia2vec\footnote{https://github.com/wikipedia2vec/wikipedia2vec} as entity embeddings. We select the \emph{ment-norm} model with K=3. We calculate the $p(e|m_i)$ using hyperlink-based mention-entity statistics from ClueWeb~\citep{Evgeniy:2013:FFA}.

\subsubsection{Results}

We discuss our results along two dimensions: (i) table-to-type linking and (ii) mention-entity linking. 
For table-to-type linking (cf. Sect.~\ref{sec:el:cs}), we
compare our method with T2K~\cite{Ritze:2015:MHT} and find that our method performs better than this baseline (0.93), reaching 0.95 
F-1 score. 
T2K leverages top-K candidates for majority vote, while we only use top-1 (cf. Sect. ~\ref{sec:el:cs}). The result indicates that most of the top-ranked candidates have the same (DBpedia) type as the tables and, thus, only keeping the top candidates for determining the table type is sufficient.
%
We encounter two types of errors for this task. 
First, mistyped by mentions, e.g., if mentions in a ``currency'' column are country aliases, we mistype them as ``country''. 
Second, a tie might exist for mentions with different types, in which case we return multiple types. 
This happens, for instance, for mentions that may have a film and a book interpretation.

For entity linking, we report results in Table~\ref{tbl:el} on two test collections. 
In the top block of Table~\ref{tbl:el}, we compare five baselines from the literature. 
We find that T2K Match outperforms DBpedia Lookup on all of the three metrics when evaluated on T2D, but performs worse than DBpedia Lookup on W2D. 
The Hybrid methods in~\citep{Efthymiou:2017:MWT} are the state-of-the-art methods. 
Except when using T2D in terms of Precision, the hybrid methods outperform T2K in both of the test collections, i.e., Hybrid I performs the best on T2D and Hybrid II performs the best on W2D among the baselines. 
The deep learning method in~\citep{Le:2018:IEL} achieves results comparable to hybrid methods.
The bottom block lists our methods: the first approach (Wikipedia API search) only uses Wikipedia and again keeps the top-10 ranked entities. 
The second (Wikipedia + Entity Disambiguation) performs entity disambiguation on top of Wikipedia search, and the last one performs all the steps in Sect.~\ref{sec:sub:el}.
We first compare Wikipedia search against T2K. 
Wikipedia search outperforms T2K on all metrics. 
T2K selects candidates only based on entity labels using Jaccard similarity, and filters out the candidates by type in the early stage, which result in a low recall. 
Wikipedia search proves that a better candidate selection is important by considering both the label and other textual fields.
When performing entity disambiguation in a later phase, the performance improves further, indicating that the table type is an effective signal for entity disambiguation.
In turn, our final method (bottom-most row) beats T2K, the hybrid methods, and the deep learning method.
T2K uses value similarity in this step, which brings in more noise because of the low relation coverage between raw tables and KBs.
The deep learning-based method in~\citep{Le:2018:IEL} optimizes the relations between mentions based on mention-entity hyperlink count statistic, while our method considers the popularity in the candidate selection step, and additionally consider lexical and semantic similarities.
%

\subsubsection{Analysis}
Comparing to the two test collections, we find that T2K tends to perform better for bigger tables with more rows. 
Our method 
performs consistently on two test collections, where W2D has smaller tables than T2D.
Next, we report on the importance of individual features for entity linking {(cf. Sect.~\ref{sec:el:cc})}, measured in terms of Gini importance. 
Both the semantic feature and and string-level similarities between the mention and candidate entity identifier are important for entity linking. 
The least important features are the type and disambiguation tag indicators. 
We additionally discuss scalability.
For candidate selection, we retrieve top-$k$ results for $l$ mentions and perform pairwise matching for candidate classification, amounting to a complexity of $O(k \times l)$.

As a conclusion, Wikipedia search can works excellently for candidate selection. 
column values are not essential for entity linking as they are quite heterogeneous and hard to normalize for raw tables. 
Lastly, table type works better in a later stage as entity disambiguation, while taking table type to exclude candidates at the beginning might result in a low coverage because of the incompleteness of KBs and type inconsistency inside tables.

\subsection{Column heading property matching}

After evaluating entity linking in Sect.~\ref{sec:eval:el}, we now turn to evaluating column heading property matching.


We train the column heading property matching model using T2Dv2 test collection.
The T2Dv2 dataset has 618 gold-standard heading-to-property correspondences for 237 tables.\footnote{http://webdatacommons.org/webtables/goldstandardV2.html} 
This dataset also includes the table-to-type gold standard.
We execute column heading matching for these linkable tables, and about 3M tables have at least one correspondence and 181,710 tables have at least two heading-property correspondences. 
For this task, we use the same evaluation metrics as for entity linking above.

\begin{table}
	\caption{Results of Column Heading Matching. The results are evaluated using T2D, which is the only publicly available heading-to-property test collection for the our table corpus.}
	\begin{tabular}{l l l l}
		\toprule
		Method & Precision & Recall & F1 \\
		\toprule
		T2K~\cite{Ritze:2015:MHT} & 0.77 & 0.65 & 0.70 \\
		Our & \textbf{0.98} & \textbf{0.71} &  \textbf{0.82} \\
		\bottomrule
	\end{tabular}
	\label{tbl:chm}
\end{table}
The results of the column heading matching are presented in Table~\ref{tbl:chm}. 
The first method is T2K Match, which is the state-of-the-art method from the literature, and the second method is our entity-assisted column matching. 
Our method outperforms T2K by 17\% regarding F1 and 29\% in terms of Precision, which is 0.98. 
We observe that T2K suffers from low recall; our method delivers a 9\% improvement against this state-of-the-art baseline.

{
Turning to the importance of individual features for this task {(cf. Sect.~\ref{sec:sub:cm})}, we find that value similarity and label similarity contribute equally.
As a result, our method beats T2K, which only utilizes the label similarity.
Besides, in contrast with entity linking, we find that column heading matching relies much more on value similarity.
}


\section{Evaluating Novel Entity Discovery}
\label{sec:ned1}

In this section, we present our experimental results for novel entity discovery.
Given the novelty of this task, no public test collections exist and we manually created (and publicly released) three data sets.

\subsection{Novel Entity Discovery Classification}

\subsubsection{Experimental Setup}

After table-to-KB matching, 
3M tables have at least have one mention-entity correspondence. 
There are 3.14M unlinked core column mentions from these linked tables. 
We sample 20k unlinked mentions and create crowdsourcing experiments to collect category labels using Figure Eight, i.e., each unlinked mention will be labeled as an in-KB entity, out-of-KB entity, or not an entity. 

\paragraph{Relevance assessments}
For novel entity classification, the annotators are asked to issue a search using the mention to get a better understanding. 
If the mention refers to something with a Wikipedia article, they are tasked to select ``in-KB entity''. 
If the mention does not refer to something with a Wikipedia article, but might be a new article---such as a private company or product that has not been included in Wikipedia yet---they are asked to select ``Novel entity'' (out-of-KB entity). 
This category aims to define the potential mentions that might be added to Wikipedia. 
The annotators are instructed to select ``Not an entity'' otherwise. 
All instances are seen by at least 3 annotators. 
The inter-rater reliability (Fleiss' kappa) is 0.4827, which is considered moderate agreement. 
The final category is decided by majority vote. 
In the end, 38.89\% of the mentions are in-KB entities, 55.22\% are out-of-KB entities and 5.90\% are not entities.

\subsubsection{Baselines} 
We consider methods in~\citep{Lin:2012:NNP} as the first baseline.
In specific, the slope and $R^2$ of the best line are each computed as a single method.
We utilize the table's contextual time stamp to find the time when it was last edited.
Additionally, we consider a feature that identifies the year when a mention first appeared in any table (``UsageSinceYear''), and another feature ``Frequency'' which indicates the total number of year-occurrence pairs.
Additionally, we consider the semantic-based entity discovery as the other baseline. It
considers three features spaces: a neural embedding space, a topical space and a lexical space~\cite{Wu:2016:EMF}.

\paragraph{Experiments and Evaluation Metrics} 
We consider the origin characteristic and salience-assisted entity discovery features in Sect.~\ref{sec:sub:ned} alone, and then combine them with semantic-based entity discovery as our final combination, namely OSS.
We train the above models and report the results using 5-fold cross-validation by using Random Forests.\footnote{We also experimented with Support Vector Regression, Gradient Boosting regression, Adapted Boost and Logistic Regression classifiers. However, we only report the method with the best performance.} 
The test collection is well-balanced and we evaluate the novel entity discovery in terms of accuracy, which is consistent with the evaluations in~\cite{Lin:2012:NNP, Wu:2016:EMF}. We further employ our novel entity discovery classification method to all unlinked mentions and find about 900k novel entities.

\subsubsection{Results}
\begin{table}
	\caption{Results of the Novel Entity Classification task in terms of accuracy. After filtering out ``naught'' entities, there are about 98\% of the mentions in the golden test collections that are either in-KB entities or out-of-KB entities. The binary classification results are listed here.}
	\begin{tabular}{l c }
		\toprule
		Method  & Accuracy \\
		\toprule
		Slope~\cite{Lin:2012:NNP} & 0.58 \\
		R square~\cite{Lin:2012:NNP} & 0.57  \\
		\citet{Lin:2012:NNP} & 0.64 \\
		\midrule
		Origin characteristic & 0.76 \\
		Semantic-based entity discovery~\cite{Wu:2016:EMF}  & 0.64 \\
		Salience-assisted entity discovery & 0.78 \\
		\midrule
		OSS & \textbf{0.83} \\
		\bottomrule
	\end{tabular}
	\label{tbl:nec}
\end{table}

We report the novel entity classification results in Table~\ref{tbl:nec}. 
The first block displays the baseline results in~\citep{Lin:2012:NNP}. 
The Slope method achieves an accuracy of 0.58, which is slightly better than $R^2$. 
When combining all the features from~\cite{Lin:2012:NNP}, we obtain a further improvement with an accuracy of 0.64. 
The second block shows the results of three methods introduced in Sect.~\ref{sec:ned}.
Semantic-based entity discovery~\citep{Wu:2016:EMF}, which is regarded as another baseline, obtains a score of 0.64.
The accuracy of the origin characteristic method is 0.76, which is an 18\% improvement over the baseline from the literature. 
The salience-assisted entity discovery method achieves an accuracy of 0.78. 
We further improve the accuracy by combining all the three methods and, in the end, obtain 0.83 accuracy (OSS). 
This result tells us that these methods complement each other for this task.
After applying our method to the unlinked mentions that appear in at least two tables in the whole WDC table corpus, we find about 900k mentions as potential novel entities.

\subsubsection{Analysis}
Next, we assess the influence of search settings, such as the field selection and the top-$k$ used, when computing the ``nearest Wikipedia distance'' in the salience-assisted entity discovery.
Table~\ref{tbl:ss} shows the accuracies of the Wikipedia distance method (cf. Eq.~\ref{eq:wd}). 
The second column shows the results of using only Wikipedia titles when searching the candidate entities and the third column shows the results of using both Wikipedia title and page content. 
Generally, using only Wikipedia titles works better than also using page content because the content tends to add more noise than signal. 
\citet{Wu:2016:EMF} report that the selection of $k$ will affect the novel entity discovery performance.
We compare the results in different settings of $k$ and find that the performance decreases as more candidates are selected, but we find $k=3$ works best when content is also considered during searching.
The latter result on $k$ is consistent with the result in~\citep{Wu:2016:EMF}, which also considers the page content for representing the knowledge base in the task of novel entity discovery. 

\begin{table}
	\caption{The influence of choosing a top-$k$ for WD.}
	\vspace*{-0.5\baselineskip}
	\begin{tabular}{l l l l l l l}
		\toprule
		$k$ & 1 & 2 & 3 & 4 & 5 & 10 \\
		\toprule
		WP Title & \textbf{0.70} & 0.68 & 0.67 & 0.67 & 0.67 & 0.66 \\
		WP Title + Content & 0.61 & 0.61 & \textbf{0.67} & 0.65 & 0.65 & 0.64 \\
		\bottomrule
	\end{tabular}
	\label{tbl:ss} 
\end{table}

\vspace*{-0.25\baselineskip}
\subsection{Entity Resolution}

\subsubsection{Experimental Setup}
For type resolution, we first compute the pairwise similarities of all tables sharing the same mention.
We find that 94.43\% of the pairs' distribution-based similarity scores lie between 0.95-1.
97.2\% of the pairs' similarity exceeds 80\%. 
We sample 1,000 table pairs with an overlapping mention in the core column following the similarity distributions; Table~\ref{tbl:sm} lists 5 example mentions. 
For mention resolution, we first sort all the unlinked mentions in alphabetical order. 
We sample 250 mentions and take their top-5 closest mentions by edit distance as candidates, e.g., for ``Zonare Medical Systems, Inc.'', the closest mentions are ``Zonare Medical Systems Inc.'', ``Zonare Medical Systems. Inc.'', ``Zonar Systems, Inc.'', ``Zonar Systems Inc.'' and ``Zonar Systems''. 
We take 5 pairs of mentions for each candidate with 1,000 pairs for annotation.

\begin{table}
	\caption{Examples of mentions for entity resolution.}
	\begin{tabular}{l l}
		\toprule
		Mention for Type Resolution & Occurrence \\
		\toprule
		IGT & 3323\\
		St.Albans & 1194\\
		Samsung Display Co., Ltd. & 708\\
		EGFR & 336 \\
		Game Boyz & 60\\
		\bottomrule
	\end{tabular}
	\label{tbl:sm} 
\end{table}
%

\paragraph{Relevance assessments}
For the entity type resolution, given a mention $m$, we resolve it by comparing pairs of tables having $m$, i.e., if the tables have different table types, we resolve $m$ with two different types. 
For surface form resolution, we provide the annotators with the pair of mentions and the tables they are from. 
They are then asked to judge if the two mentions refer to the same entity by comparing the table content and search results for the mentions.

\paragraph{Experiments and evaluation Metrics} 
We train Mention2vec using Word2vec for all the linkable tables, which is used for computing the embedding similarity of two mentions. 
We use the same form of cross-validation as above and find the optimal the threshold to decide if two mentions refer to the same entity as being 0.95. 
Then, four string-level similarities and table-table similarity using different fields are computed as features. 
We combine the string and table features to train a model.
We again report the results using the same form of cross-validation as above.

\begin{table}
	\caption{Results of Entity resolution, where the bottom block lists the surface form results.}
	\begin{tabular}{l l l l l}
		\toprule
		Task & Accuracy & Precision & Recall & F-1 \\
		\toprule 
		Type resolution & \textbf{0.99} & \textbf{0.99} & \textbf{0.99} &  \textbf{0.99} \\
		\midrule 
		Mention2vec & 0.84 & 0.86 &  0.97 & 0.91 \\
		Table similarity  & 0.83 & 0.93 &  0.88 &  0.91 \\
		String similarity  & 0.94 & 0.97 & 0.96 & 0.97 \\
		Table + String  & \textbf{0.96} & \textbf{0.99} & \textbf{0.96} & \textbf{0.97} \\ 
		\bottomrule
	\end{tabular}
	\label{tbl:er}
\end{table}

\subsubsection{Results and Analysis}
We discuss entity resolution by evaluating the type resolution and surface form resolution, and report the results in Table~\ref{tbl:er}.

\paragraph{Type Resolution}
We report results on distribution-based similarity as the accuracy of the type resolution task is 0.99 using 5-fold cross-validation, i.e., we only fail to resolve 12 pairs out of 1,000 instances. 
The precision, recall and F-1 are all exceeding 0.99.
We also conduct an error analysis for this task.
For type resolution, most of the 12 incorrectly resolved instances are people, e.g., for a football player ``Frank'' in two athlete tables, we mistake them for being the same person given that they belong to the same ``athlete'' type, whilst in fact they are two different individuals. 
We will require additional information to resolve such cases and will leave this for future work.

\begin{figure}[t]
	\centering
	\includegraphics[scale=0.5]{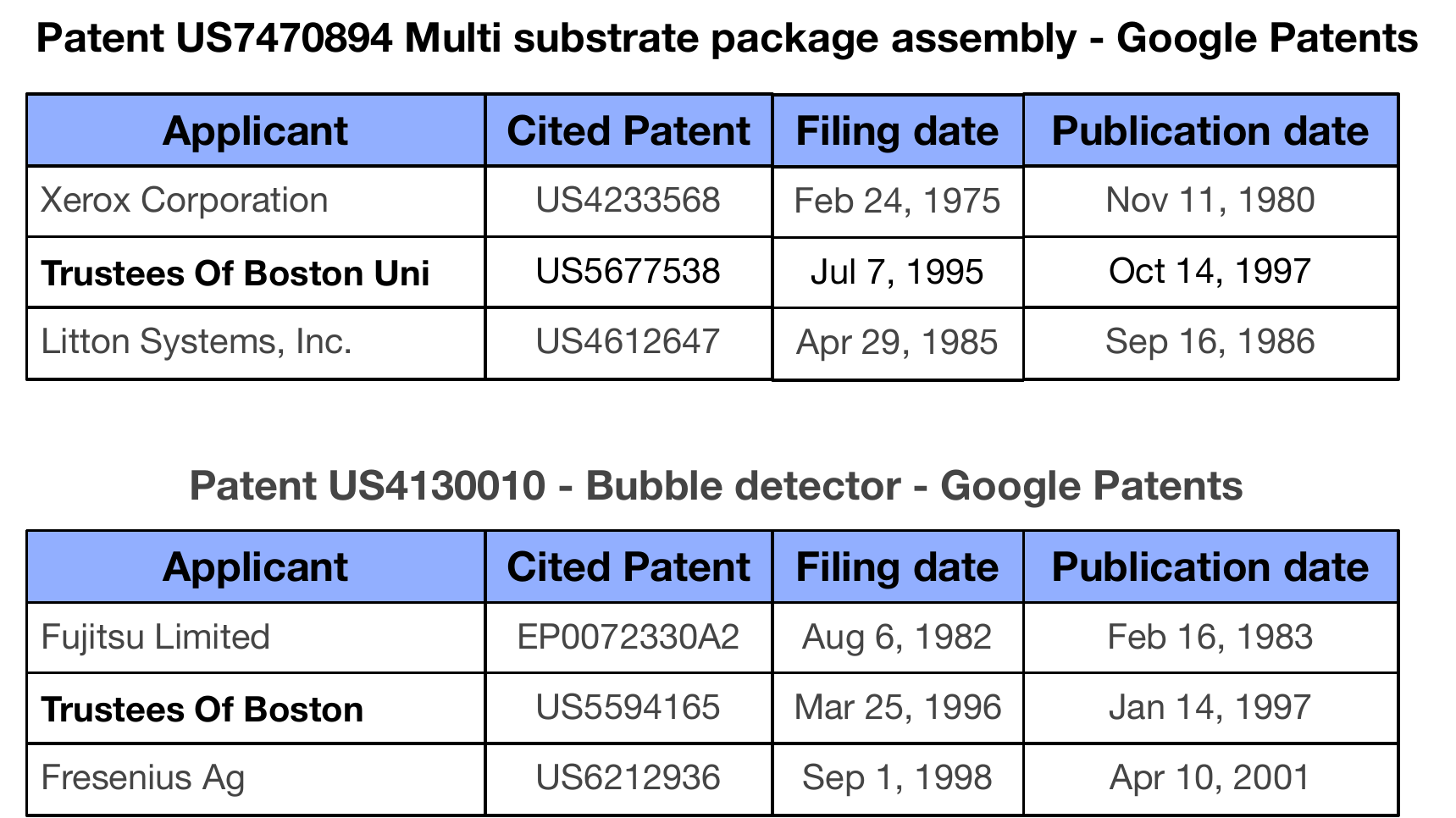}
	\caption{Examples of incorrectly resolved mentions in two similar tables. These two tables have the same set of headings, similar captions, and the same type of linked entities. }
	\label{fig:ana}
\end{figure}

\paragraph{Surface Form Resolution}
We report the surface form resolution results in the second block of Table~\ref{tbl:er}.\footnote{Only the best results are reported here after experimenting with multiple machine learning classifiers.} 
The Mention2vec method achieves an accuracy of 0.84 and the table similarity of 0.83. 
String similarity has the best performance among the single-component methods with an accuracy of 0.94. 
We further combine the String and Table methods as our final method and achieve an accuracy of 0.96. 
The precision of the combined method is 0.99 and the recall reaches 0.97, which are both satisfactory results.
For surface form resolution, the instances we incorrectly resolved all have very similar expressions and types.
For example, ``Trustees Of Boston University'' and ``Trustees Of Boston'' have very similar expressions, but they refer to trustees of different entities.
Also, they have the same type and very similar table content. 
See Fig.~\ref{fig:ana} for an illustration.
We determine feature importance based on Gini scores for the Table + String method and find that the two most important features are Jaccard similarity and edit distance between labels, followed by the distribution-based method. 
The least important feature is the Jaccard similarity between two tables' headings.
Generally, the string-level similarity and the type distribution-based cosine similarity methods work better than the other ones.

\section{Conclusion}

We have introduced the task of novel entity discovery from Web tables, which aims to find new knowledge from relational tables on the Web for populating a knowledge base. 
To this end, we defined two tasks: table-to-KB matching and novel entity discovery. 
To keep track of the performance of each component, we address them in a pipeline architecture instead of an end-to-end fashion.

The first task, table-to-KB matching, aims to match a table with elements of the KB and includes entity linking and column heading matching.
We have employed a feature-based algorithm for entity linking and entity-assisted column heading matching method across the whole table corpus and show that our methods achieve large relative improvements over the state-of-the-art baselines on public test collections, especially in terms of precision whilst also improving recall.
%
We find that both lexical and semantic features contribute to candidate classification and, additionally, that the table type on its own provides a strong signal for entity disambiguation.

The second task, entity discovery, involves novel entity classification and entity resolution and we have evaluated our approaches by using three novel test collections custom-built using crowdsourcing. 
For this task we also find that we considerably improve precision whilst keeping recall stable. 
The three main features (origin characteristic, saliency, and semantic similarity) complement each other and we identified the table type as an important indicator for constituent entity type resolution. 

For future work, we aim to use the developed methods and table-to-KB correspondences for other table-related applications such as novel schema discovery and inference.

\bibliographystyle{ACM-Reference-Format}
\bibliography{00paper}

\end{document}